\RequirePackage{fix-cm}
\documentclass[twocolumn]{svjour3}          % twocolumn
\smartqed  % flush right qed marks, e.g. at end of proof
\usepackage{graphicx}
\usepackage{cite}
\usepackage{amsmath}

\begin{document}

\title{Simple correlated wave-functions for excitons in 0D, quasi-1D and quasi-2D quantum dots\thanks{Support from MINECO project CTQ2014-60178-P, UJI project P1-1B2014-24 is acknowledged.}
}


\author{ Josep Planelles}


\institute{Josep Planelles \at
              Universitat Jaume I \\
			  Av. de Vicent Sos Baynat, s/n \\
			  12006 Castell\'o de la Plana, (Spain)\\
              Tel.: +34 964 728090\\
              Fax: +34 964 728066\\
              \email{josep.planelles@uji.es}           %  \\
}

\date{Received: date / Accepted: date}

\maketitle

\begin{abstract}
We propose correlated yet extremely simple single-parameter-dependent wave-functions with a Slater-type correlation factor, to describe excitons in 0D, quasi-1D and quasi-2D semiconductor quantum dots. We provide closed-form formulas for the wave-function normalization factor, electron/hole single-particle density and the expectation value of the kinetic energy. We additionally supply fast integration procedures for the Coulomb interaction in the presence of dielectric mismatch with the surrounding medium for nanoplatelets (quasi-2D systems), and for the bare-Coulomb integral in long nanorods (quasi-1D systems).
\end{abstract}

\vspace{0.5cm}
%%%%%%%%%%%%%%%%%%%%%%%%%%%%%%%%%%%%%%%%%%%%%%%%%%%%%%%%%%%%%%%%%%%%%%%%%%
\noindent {\it Se me ha muerto como del rayo Claudio-Zicovich, con quien tanto queria.} 
%%%%%%%%%%%%%%%%%%%%%%%%%%%%%%%%%%%%%%%%%%%%%%%%%%%%%%%%%%%%%%%%%%%%%%%%%%
\section{Introduction}
\label{intro}
Semiconductor colloidal quantum dots (CQDs) have been synthesized for more than 30 years now, and their synthesis is becoming mature enough that these nanoparticles have started to be incorporated into devices.\cite{Lhuillier, Chhowalla} The control of the shape (cuboid,\cite{Protesescu} plates,\cite{Lhuillier, Akkerman, Ithurria} rods,\cite{Peng} wires,\cite{Imran, Yu}) of CQDs brings a unique way to tune the confinement, 0D, quasi-1D, or quasi-2D, of the charge carriers and as a consequence their density of states.

\noindent The key problem in the investigation of electronic and optical properties of QD’s is finding the energy spectrum of confined charge carriers and the corresponding wave functions. The Coulomb interaction between the conduction band electron and the valence band hole in the exciton influences the energy of optical absorption and photoluminescence. The dielectric mismatch at the interfaces of CQDs has also a considerable effect on the exciton energies, the dielectric enhancement of excitons  being demonstrated in quantum wells, wires and dots. \cite{Fonoberov}

\noindent The usual approach for obtaining the energy and eigenfunctions is based on the variational property of the expectation values of the energy. The most employed variational methodology first determines self-consistently the single particle orbitals (mean field theory) and then, the correlation energy is accounted for by means of a configuration interaction (CI) expansion. \cite{Szabo}

\noindent The nature of electron-hole correlation is, though, very different from electron-electron correlation typically encountered for the ground state calculations in many-electron systems because the particles involved are oppositely charged. As a consequence of the attractive Coulomb interaction, the quality of the electron-hole wave function at small inter-particle distances becomes very important. This has relevant consequences in particular on the calculation of electron-hole (eh) recombination probability $P_{eh}$. Since $P_{eh}$ is the overlap of presence probabilities of electrons and holes, an accurate description of the electron-hole wave function at small electron-hole distances is extremely important.

\noindent It is worth to draw attention to the fact that there are systems with genuine (bare-) Coulomb interactions which are exactly solvable. For example, two interacting particles in an external harmonic-oscillator potential.\cite{Taut, Jacek1, Jacek2, Jacek3} Independently of the development of highly efficient and sophisticated numerical approaches for solving Schr\"odinger equation, a search for new exact and quasiexact solutions always was and still is of interest from both practical and formal perspectives.

\noindent However, accounting for realistic spatial and dielectric confinement and Coulomb polarization in CQDs requires more sophisticated methods. As pointed out above, CI expansions for interacting particles of oppositely char\-ged is slowly convergent. Departing from the usual CI expansion build out of self-consistent single particle orbitals can be very fruitful, allowing us to write very compact and at the same time very accurate wave functions. However it is computationally much more demanding and for this reason special care must be given to the design of an efficient method of generating and optimizing the trial wave-function. The standard way is to optimize a trial wave-function using Variational Monte Carlo VMC, either minimizing the local energy or the variance of the local energy. A popular and effective approach to building compact explicitly correlated wave-functions for many-electron systems is to multiply a determinantal wave-function by a correlation factor, the most commonly used being a Jastrow factor.\cite{Jastrow} The inclusion of the Jastrow factor does not allow the analytical evaluation of the integrals, so the use of VMC is usually mandatory. The form of the wave function as a product of a sum of determinants and a generalized Jastrow factor relies on the idea that near-degeneracy correlation is most effectively described by a linear combination of low-lying determinants whereas dynamic correlation is well described by the generalized Jastrow factor. With this kind of wave-functions Filippi and Umrigar, using diffusion Monte Carlo, were able to recover 93\% -- 99\% of the correlation energy for a set of first-row homo-nuclear diatomic molecules.\cite{Filippi}

\noindent To assure a high-quality wave-function it is particularly important that the wave-function satisfy the cusp conditions,\cite{Kato,Pack}  representing the behavior of the exact wave function at the coalescence of two particles. It is also important to take into account the asymptotic conditions,\cite{Patil} which represent the behavior when one of the particles goes to infinity. Bertini et al.\cite{Bertini} have employed explicitly correlated trial wave-functions for ground and excited states of Be and Be$^{-}$ fulfilling cusp and asymptotic conditions by means a Pad\'e factor $\exp[(a r + b r^2)/(1 + c r)]$ for the electron--nucleus part  and a Jastrow factor $ \exp[ a r / (1 + c r)]$ for the inter-electronic part. The Pad\'e factor is a good choice for the electron--nucleus part, because it is the best compromise between flexibility and small number of parameters. In fact this function goes as $\exp[a r]$ for $r \to 0$ and $\exp[(b/c)r]$ for $r \to \infty$, so with different exponents it can accommodate both the coalescence at the nucleus and the decay for large r.

\noindent Although the precise form of the correlation factor is not as important in highly accurate computations of small systems, appropriate correlation factors are essential to approach chemical accuracy in modest basis sets. It appears that a single Slater-type geminal factor $\exp[a \, r_{12}]$ is very close to optimal.\cite{Klopper1} However, the use of Gaussian-type geminal (GTG) has proved also to be successful in correlated configuration interaction.\cite{Elward} The principle reason for using GTGs as opposed to Slater-type  or Jastrow functions is that integrals involving GTGs are known analytically\cite{Persson} and are much faster to compute than integrals involving Slater and Jastrow functions. Obviously, for calculations involving GTGs, the correlation function is not  just a Gaussian function. In this case, the one-electron basis should be improved iteratively by adding GTGs with increasing angular momentum quantum number.\cite{Klopper2}

\noindent As pointed out by Patil and Tang,\cite{Patil_book} often the complexity of accurate complex variational wave-functions does not allow a transparent, compact description of the physical structure. In these cases, the global and local properties of wave functions can provide deeper insight, useful guidelines and criteria in the development of accurate and compact wave functions of many-particle systems. In this sense, Patil\cite{Patil_epjd} has been able to draw simple parameter-free wave-functions  for two- and three-electron atoms and ions yielding fairly accurate values for the energies, $\langle r^{2n}\rangle$, multipolar polarizabilities of two-electron atoms and ions, and for the coefficients of the asymptotic density. He has also obtained simple wave-functions  for the lowest energy state and first excited state of a confined hydrogen atom just relaying on a simple coalescence property near the center, and an inflexion property at the boundary and nodal points, the predictions for the energies and multipolar polarizabilities calculated being in close agreement with accurate, numerically obtained, values.\cite{Patil_jpb}

\noindent All the same, Prendergast et al.\cite{Prendergast} explored  the effect of the electron cusp on the convergence of the energy for CI wave functions and concluded that the description of the electron cusp as such is not a limiting factor in calculating correlation effects with configuration interaction methods. In their study, they introduced a fictitious electron-–electron interaction which, unlike the true Coulomb potential, does not diverge at electron coalescences and therefore has many-body eigenfunctions which are smooth there. The replacement of the divergent Coulomb interaction with a finite interaction leaves the convergence properties largely unchanged. Then, contrary to what is often stated in the literature (that failure of the CI expansion to reproduce the correct electron cusp, i.e., the short-range part of the Coulomb hole, is what leads to a slow convergence in the energy with respect to the number of configurations) they attributed to medium-range correlations, which are present for both types of electron interaction, the reason of the slow convergence of the CI expansion.

\noindent  It should be finally mentioned that simple variational models as that by Romestain and Fishman,\cite{Romestain} is able to describe a bare-Coulomb interacting exciton in a cubic 0D quantum dot using a Slater type of correlated functions and involves at most 3D integrals. Also we may quote earlier models, also restricted to bare-Coulomb interacting Hydrogenic impurity states\cite{Garnett1} and excitons in quantum boxes.\cite{Garnett2} These models employs linear combination of Gaussian functions of the electron-hole separation as a correlation factor and numerically integrates the Coulomb interaction by means a Fourier representation of the Coulomb potential. 

\noindent The aim of this paper is to introduce correlated yet extremely simple single-parameter-dependent wave functions for excitons in 0D, quasi-1D and quasi-2D CQDs. We derive closed-form formulas for the normalization factor, single-particle electron/hole density and the expectation value of the kinetic energy. Exactly solvable models are scarce and valuable as can be used to improve approximate methods. For example, Kais et al.\cite{Kais} employed the analytically solvable version of the Harmonium, i.e. a two-electron atom, in which the elec\-tron-electron repulsion is Coulombic but the electron-nucleus attraction is replaced by a harmonic oscillator potential, to obtain, by inversion procedure, the exchange and correlation Kohn-Sham functionals. We provide here single-particle electron/hole density of excitons, i.e. two distinguishable particles with opposite charges and different isotropic or anisotropic masses confined in 0D, quasi-1D and quasi-2D quantum dots that can be useful just to carry out calculations on excitons and also to improve approximated methods for confined multi-excitons which are of increasing interest for a variety of high-fluence optoelectronic applications, including photovoltaic devices where low-threshold laser gain and ultrafast energy transfer are desirable.\cite{RowlandNM,KlimovNAT} We additionally provide fast integration procedures for the Coulomb interaction in typical quasi-2D nanoplatelets including polarization of the Coulomb interaction produced by the dielectric mismatch between the quantum dot and the surrounding medium, and for the bare-Coulomb integral in quasi-1D long nanorods. For nano-platelets, the original sixfold integral is reduced to some analytical and a twofold numerical integral, while a one-coordinate numerical integration is required for the bare Coulomb interaction in quasi-1D long nanorods.

\noindent The paper is organized as follows. Section II presents an illustrative 2D model calculation to show the different performance of Hartree-CI vs. functions with a variational correlated factor (Slater,  Gaussian and mixed). In section III we present the models for nano-platelets (quasi-2D), long nanorods (quasi-1D) and cuboid (0D) CQDs. The paper ends with several appendixes with mathematical details. 

\section{2D model calculation}
\label{sec:2}

In this section we enclose a set of illustrative calculations on a Hamiltonian model analytically solvable to show the low convergence of the Hartree-CI vs. the good performance of  simple correlated functions. The system is a 2D exciton in a uniform dielectric media confined by a Harmonic potential,
\small
\begin{equation}
\label{eqII.1}
\hat H = \frac{\hat p_e^2}{2 m_e}+\frac{\hat p_h^2}{2 m_h}+\frac{1}{2} \omega^2 (m_e r_e^2+m_h r_h^2)-
          \frac{1}{\epsilon |{\mathbf r}_e-{\mathbf r}_h|}
\end{equation}
\normalsize
\noindent In terms of center of mass and relative motion, $\hat H =\hat H_R +\hat H_r $ with,
\small
\begin{equation}
\label{eqII.2}
\begin{array}{lll}
\hat H_R &=& \frac{\hat P^2}{2 M}+\frac{1}{2} M \omega^2 R^2\\
\\
\hat H_r &=& \frac{\hat p^2}{2 \mu}+\frac{1}{2} \mu \omega^2 r^2 -\frac{1}{\epsilon r}
\end{array}
\end{equation}
\normalsize
\noindent where $M$ is the total and $\mu$ the reduced mass, and $\mathbf R$ and $\mathbf r$ are the center of mass and relative motion coordinates. $\hat H_R$ is the 2D Harmonic oscillator which ground state energy and wave-function are well known: $E_R=\hbar \omega$ and $\Psi(R) = \frac{\beta}{\sqrt{\pi}} \exp[-\beta^2 R^2/2]$ with $\beta=\sqrt{M\omega/\hbar}$. $\hat H_r$ is the harmonically confined 2D hydrogen atom that has energy and wave-functions in closed-form expressions for some particular confinements\cite{Koscik} and that we numerically solve up to the desired accuracy.

\noindent We carry a CI expansion of Hamiltonian (\ref{eqII.1}) including for both, electron and hole, up to three $m=0$, two $m=\pm 1$ and one $m=\pm 2$  orbitals, where $m$ is the angular momentum quantum number. In order to get the numerical orbitals for the CI calculation we first carry out numerical self-consistent (SCF) calculations in cylindrical coordinates for the different angular momentum quantum number $m$ for electron in the presence of a ground $m=0$ hole (and for a hole in the presence of a ground $m=0$ electron). In the SCF calculation we write the $1/r_{eh}$ in terms of Bessel functions,\cite{Note1} 

\small
\begin{flalign} 
\label{eqII.3}
& \frac{1}{\sqrt{r_e^2+r_h^2 -2\, r_e r_h \cos (\phi_e-\phi_h)}} = & \nonumber \\
&\hspace{0.75cm} =  \sum_{m=-\infty}^{m=\infty} e^{i m (\phi_e-\phi_h)}\; \int_0^{\infty} J_m(k \, r_e) J_m(k \,r_h)dk &
\end{flalign} 
\normalsize

\noindent In particular, when dealing with integrals whose orbitals have angular momentum quantum numbers differing by $\Delta m=0$, we employ the following identity: 
\small
\begin{equation}
\label{eqII.4}
\int_0^{\infty} J_0(k r_i) J_0(k r_j) dk =\frac{2 K(r_j^2/r_i^2)}{\pi r_i}
\end{equation}
\normalsize
\noindent where $r_i > r_j$ represent $r_e$ and $r_h$ and $K(x)$ is the complete elliptic integral of the first kind.  In a similar way, an integral involving orbitals with angular momentum quantum numbers differing in $\Delta m=a$, then we get proper combination of elliptic functions corresponding to the  integration $\int_0^{\infty} J_a(k r_i) J_a(k r_j) dk$.

\noindent An alternative approach is the use of the center-of-mass and relative motion form of the Hamiltonian, eq. (\ref{eqII.2}). We may initially disregard the harmonic confinement in the relative motion and treat it later as a perturbation. The center-of-mass, as stated above, is a 2D harmonic oscillator which ground state energy is 
$E_R=\hbar \omega$, and the relative motion is just the 2D hydrogen atom, with energy $-\frac{\mu}{\epsilon^2} \frac{1}{2(n-1/2)^2}, \; n=1,2,\dots $ a.u.\cite{Yang} and a wave-function properly accounting for the electron-hole coalescence.  Next, we consider the harmonic confinement as a perturbation and calculate the energy of the confined exciton either just as a perturbation (i.e. calculating the expectation value of the harmonic confinement) or carrying out linear variations with the eigenfunctions of the 2D hydrogen atom (for the sake of symmetry only $s$-type of orbitals are involved in this calculation). Explicit formulas for the expectation values $\langle n,\ell|r^2|n,\ell\rangle$ can be found in the paper by Yang et al.\cite{Yang}. The rest of needed integrals can be obtained in closed form. For example, $\langle 1,0|r^2|2,0\rangle =-\frac{27 \sqrt{3} \epsilon^2}{128 \mu^2}$ a.u. 

\noindent Another even simpler approach is the inclusion of a variational parameter in the Slater-type ground state eigenfunction of the 2D Hydrogen-like atom, i.e., consider the normalized wave-function $R(r)= 2 a \exp[-a r]$, with $a=2 \alpha \mu/\epsilon$ and $\alpha$ the variational parameter. An elementary calculation yields the relationship between the confinement frequency $\omega$ and the optimal $a$ value (and therefore the relationship with the optimized variational parameter $\alpha$). In a.u. it reads: 
\small
\begin{equation}
\label{eqII.5}
\omega=\frac{1}{\mu}\sqrt{\frac{2}{3}}\sqrt{a^4-2 \,\frac{\mu}{\epsilon}\, a^3}
\end{equation}
\normalsize
\noindent We can also employ a normalized Gaussian-type variational function   $R(r)= 2 \sqrt{b} \exp[-b r^2]$. In this case the relationship between the confinement frequency $\omega$ and the optimal $b$ is:
\small
\begin{equation}
\label{eqII.5b}
\omega=\sqrt{\frac{2}{\mu}} \, \sqrt{\frac{2 b^2}{\mu}-\frac{b^{3/2}\, \sqrt{2 \pi}}{\epsilon}}
\end{equation}
\normalsize
\noindent Finally, we can employ a two parameters variational function $R(r) = N \exp[-a r -b r^2]$, with $N$ the normalization factor,  that incorporates the two limit cases: the free exciton, described by the Slater function and the extremely strong confinement limit, where the Coulomb interaction is disregarded yielding a 2D harmonic oscilla\-tor-like described by a Gaussian function.

\begin{figure}
\includegraphics[width=0.5\textwidth]{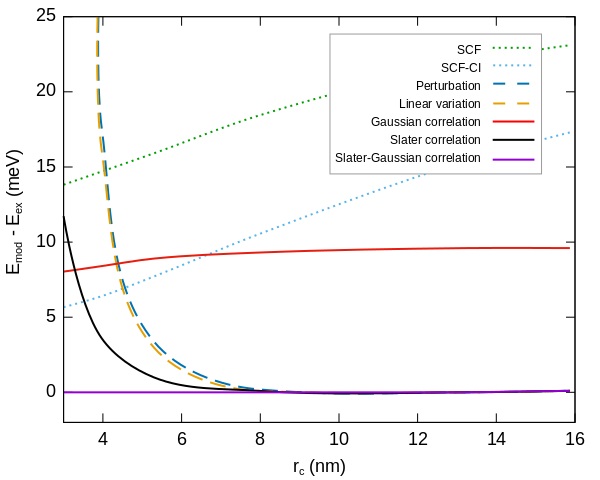} 
\caption{Energies, relative to the exact $E_{mod}-E_{ex}$, vs. confinement radius $r_c$, for the different approximate models of a two-dimensional exciton harmonically confined.}
\label{fig:1}
\end{figure}

\noindent In figure 1 we collect the results of the different models particularized for CdSe, a typical semiconductor. In this case, the electron effective mass is isotropic, $m_e=0.12$, while the hole is highly anisotropic. Then, we should take the heavy hole in-plane effective mass $m_h= \frac{1}{\gamma_1+\gamma_2} \approx 0.15$. Finally, the dielectric constant is $\epsilon =9$.\cite{Woggon} In the figure we represent the difference between the energy obtained by the different models and the exact energy vs. the confinement strength represented by the 2D harmonic oscillator confinement radius $r_c$. The confinement radius $r_c$ is related to the confining frequency $\omega$ by $\omega=\frac{2 \hbar}{m r_c^2}$. Since electron and hole have different masses,  in the above formula we consider $m$ to be the average $m=(m_e+m_h)/2=0.135$.

\noindent Figure 1 neatly shows the different performance of the various approaches. The more sophisticated two parameter wave-function, that
almost become a Slater function ($b\to 0$) in the low-confinement regime while has a larger contribution of the Gauss part in the strong confinement regime, yields an energy indistinguishable from the exact one in all confinement regimes. The one parameter Slater function does the same, except in the very strong confinement. Interestingly, the one parameter Gaussian function departs about 10 meV from the exact, independently of the confinement regime. It may be related to the fact that this particular system becomes a 2D harmonic oscillator if we remove the Coulomb term i.e., the Gaussian-like function is most suitable to describe the limit of highly strong confinement where the Coulomb contribution is negligible.  The perturbation approach on the 2D hydrogen ground state eigenfunction has a performance similar to the one parameter Slater one, but deteriorates as the confinement get stronger. The addition of several s-orbitals to carry out a linear variation does not improve significantly the perturbation result. Finally, we can see that the (lower) accuracy of both SCF and  SCF-CI changes depending on the confinement strength, showing then the poorest behavior amongst the different studied approaches. 

\noindent The excellent behavior of the one parameter Slater function up to ~5nm, which is well below typical nanoplatelets lateral dimensions,\cite{Woggon} suggests it as the most suitable model for extensive yet reliable calculations. For this reason, in the following sections, all our models for excitons in 0D, quasi-1D and quasi-2D semiconductor quantum dots contain this correlation factor.

\section{Models for quasi-2D, quasi-1D and 0D quantum dots}
\label{sec:3}
We present in Fig.2 the three systems we deal with. Building up appropriate simple one parameter variational models is the core of this paper. The goal is to reach models with simple analytical expressions for the kinetic energy (accounting for anisotropic mass) the normalization factor of the wave-function and the one-particle density (so that any one-body potential term may be calculated by means at most a 3D volume integral). In the following subsections we detail the different models for quasi-2D, quasi-1D and 0D quantum dots. 

\begin{figure}[h]
\includegraphics[width=0.5\textwidth]{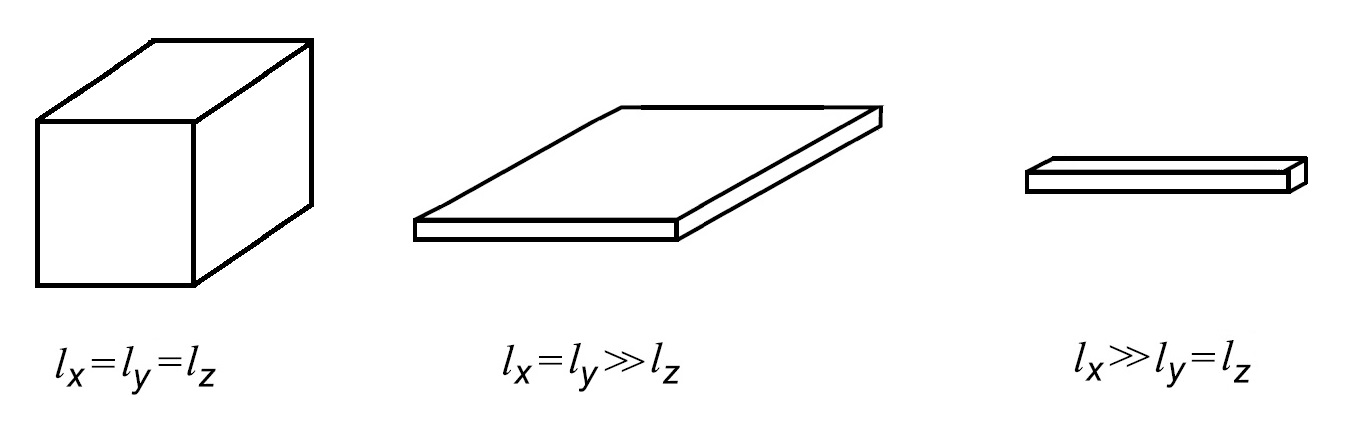} 
\caption{0D, quasi-2D and quasi-1D quantum dots}
\label{fig2}
\end{figure}

\subsection{Nanoplatelets in dielectric media: a quasi-2D system}
\label{subsec:3.1}
The Hamiltonian of an exciton in a nanoplatelet reads,
\small
\begin{flalign} 
\label{eqIII.1}
& \hat H = \sum_{e,h} \left[ -\frac{\hbar^2}{2 m_{\parallel,i}} \left( \frac{\partial^2}{\partial x^2}+\frac{\partial^2}{\partial y^2}\right)-\frac{\hbar^2}{2 m_{z,i}}\frac{\partial^2}{\partial z^2}  +V_i \right] & \nonumber \\
 &\hspace{0.7cm} + \;V_c(r_e,r_h)&
\end{flalign} 
\normalsize
\noindent where $V_i$ represent the sum of all possible single-particle potentials affecting particle $i$ and $V_c(r_e,r_h)$ represents the Coulomb interaction that may or not include dielectric effects. An infinite barrier is used to confine the exciton in the integration box. Please, note however that it does not mean that an infinite barrier is used to confine the exciton in the box, as the single particle  $V_i$ potential may have a profile that mimics the band off-set between neighboring materials. At the end of the external material, where the wave function is null, we enclose the infinite wall, i.e., we assume the wave-function to be mathematically zero.

\noindent The exciton variational wave function is chosen to be a product of the electron and hole lowest-energy subband states and a Slater correlation factor. 
\small
\begin{flalign} 
\label{eqIII.2}
& \Psi(r_e, r_h)= N \cos k x_e \cos k x_h \cos k y_e \cos k y_h \cos \bar{k} z_e  *& \nonumber \\ 
 &\hspace{2.5cm} * \cos \bar{k} z_h \; e^{-a \sqrt{(x_e-x_h)^2+(y_e-y_h)^2} \,} &
\end{flalign} 
\normalsize
\noindent where $N$ is the normalization factor, $k= \pi/L_x = \pi/L_y$ and $\bar{k} = \pi/L_z$. The form chosen for the wave-function gives the correct results for the ground state of the exciton in the limits of extremely high (small L) and negligible (large L) confinement. In the small-L limit the correlation factor becomes unity while in the large-L limit the correlation factor is the (2D) bulk-exciton ground-state wave function, and product of the electron and hole lowest-energy subband states are envelope functions which are slowly varying on the scale of the exciton.  This wave-function is similar to that employed by Bryant\cite{Garnett2} but have the advantage of using a Slater instead of a combination of Gaussian to mimic a Slater. We also provide a simple integration of the sixfold Coulomb integral, even in the presence of dielectric mismatch, only involving a twofold numerical integration. 

\noindent The use of a Slater correlation factor has the additional advantage of being the unique correlation factor having a simple additive closed-form for the kinetic energy (see Appendix A for details):
\small
\begin{equation}
\label{eqIII.3}
\langle \Psi(r_e,r_h)| \hat T|  \Psi(r_e,r_h)\rangle = \frac{\bar{k}^2}{2\mu_z}+\frac{k^2}{\mu_{\parallel}}+\frac{a^2}{2\mu_{\parallel}}
\end{equation}
\normalsize
\noindent with $k= \pi/L_x = \pi/L_y$ and $\bar{k} = \pi/L_z$ and $a$ the variational parameter to be optimized.

\subsubsection{The single-particle density}
\label{subsubsec:3.1.1}
In this section we employ the labels 1 and 2 to refer to either particle and $L$ to refer to $L_x=L_y$. The single-particle density of particle "1" reads,

\small
\begin{flalign} 
\label{eqIII.4a}
& \rho(r_1) = \int |\Psi(r_1,r_2)|^2 d^3r_2 & \nonumber \\
& = N^2 \; \frac{L_z}{2} \cos^2 \bar{k} z_1 \cos^2 k x_1 \cos^2 k y_1 * & \nonumber \\
           & * \iint\displaylimits_{-L/2}^{\;\;L/2} \cos^2 k x_2 \cos^2 k y_2 \; e^{-2 a \,\sqrt{(x_2-x_1)^2+(y_2-y_1)^2}} d x_2 d y_2 & 
\end{flalign}
\normalsize

\noindent After a rather long procedure, detailed in Appendix B, one can obtain  the following closed-form for the single-particle density $\rho(r_1)$:

\small
\begin{flalign} 
\label{eqIIIa.16}
& \rho(r_1) = N^2 \;  \frac{\pi \, L_z}{16} \; \cos^2 \bar{k} z_1 \; \cos^2 kx_1 \; \cos^2 k y_1 \;\;* & \nonumber \\
 & \hspace{1cm} *  \left[ \frac{1}{a^2}+\frac{a}{(a^2+k^2)^{3/2}} \left(\cos 2 k x_1 + \cos 2 k y_1 \right) + \right.& \nonumber \\
 & \hspace{1cm} + \left. \frac{a}{(a^2+2 k^2)^{3/2}} \; \cos 2k x_1 \cos 2 k y_1 \right] &
\end{flalign} 
\normalsize

\noindent The normalization factor, $N^2$ can now be obtained from the identity,
\small
\begin{equation}
\label{eqIII.17}
\int_{-L/2}^{L/2} \int_{-L/2}^{L/2} \int_{-L_z/2}^{L_z/2} \rho(x_1, y_1, z_1) d x_1 d y_1 d z_1 = 1 
\end{equation}
\normalsize
\noindent Taking into account that $k=\pi/L$ and $\bar{k}=\pi/L_z$, we get:
\small
\begin{equation}
\label{eqIII.18a}
N^2 =\frac{128}{\pi \, L_z^2 \, L^2} \; \left(\frac{1}{a^2}+\frac{a}{(a^2+k^2)^{3/2}}+\frac{1}{4}   \frac{a}{(a^2+2 k^2)^{3/2}} \right)^{-1}
\end{equation}
\normalsize

\subsubsection{Nanoplatelets with $L_x \neq L_y >> L_z$}
\label{subsubsec:3.1.2}
For the sake of completeness we also enclose the case where the two in-plane dimensions are not alike. The calculation is similar to that with $L_x=L_y$. The kinetic 
energy is now:
\small
\begin{equation}
\label{eqIII.18}
\langle \Psi(r_e,r_h)| \hat T|  \Psi(r_e,r_h)\rangle =\frac{k_x^2}{2 \mu_{\parallel}}+\frac{k_y^2}{2 \mu_{\parallel}}+ \frac{k_z^2}{2\mu_z}+\frac{a^2}{2\mu_{\parallel}}
\end{equation}
\normalsize
\noindent with $k_x= \pi/L_x$, $k_y= \pi/L_y$,  $k_z = \pi/L_z$ and $a$ the variational parameter to be optimized.\\

\noindent The density reads:
\small
\begin{flalign} 
\label{eqIII.19}
& \rho(r_1) =  N^2 \;  \frac{\pi \, L_z}{16} \; \cos^2 k_z z_1 \; \cos^2 k_x x_1 \; \cos^2 k_y y_1 \; * &\nonumber \\
 \nonumber \\
 & * \left[ \frac{1}{a^2}+\frac{a}{(a^2+k_x^2)^{3/2}} \cos 2 k_x x_1   + \frac{a}{(a^2+k_y^2)^{3/2}} \cos 2 k_y y_1 \right. & \nonumber   \\
 &  \left.  \hspace{0.75cm} + \;\; \frac{a}{(a^2+k_x^2+ k_y^2)^{3/2}} \; \cos 2k_x x_1 \cos 2 k_y y_1 \right] &
\end{flalign} 
\normalsize

\noindent with

\small
\begin{flalign} 
\label{eqIII.20}
& N^2 =  \frac{128}{\pi \; L_z^2 \, L_x \, L_y} \; \left[ \frac{1}{a^2}+\frac{1}{2}\frac{a}{(a^2+k_x^2)^{3/2}}+
          \right. & \nonumber \\
&	\hspace{0.75cm} \left. + \; \frac{1}{2} \frac{a}{(a^2+k_y^2)^{3/2}}  + \;  \frac{1}{4} \frac{a}{(a^2+k_x^2+ k_y^2)^{3/2}} \right]^{-1} &
\end{flalign} 
\normalsize

\noindent It should be pointed out that the dimensions $L_x$ and $L_y$ should not be very different for a good performance of the model. If it was the case, in order to keep the accuracy, the correlation factor should be supplied with an additional variational parameter. Namely, $f(r_{eh})=\exp[-a\, \sqrt{(x_e-x_h)^2+ b\, (y_e-y_h)^2}]$. A similar correlation factor has been proposed by Khramtsov et al.\cite{Khramtsov} for the case of a 0D cuboid QD.  However, the two-parameters model has not the simplicity of the single-parameter one so that no closed formulas for kinetic energy, density and normalization factors can be found. Then, the calculation becomes much heavier, unsuited for extensive calculations. 

\subsubsection{The polarized-Coulomb integral}
\label{subsubsec:3.1.3}
We deal with the integral,
\small
\begin{equation}
\label{eqIII.21}
V_c=\int |\Psi({\bf r}_e,{\bf r}_h)|^2 H_c \; d^3{\bf r}_e d^3 {\bf r}_h
\end{equation}
\normalsize
\noindent with $\Psi({\bf r}_e,{\bf r}_h)$ given in eq. (\ref{eqIII.2}) and $H_c$ represents the coulomb operator including the image charges originated in the dielectric mismatch of the nanoplatelet and the surrounding medium.\cite{Takagahara} 
\small
\begin{equation}
\label{eqIII.22}
H_c = - \sum_{n=-\infty}^{n=\infty}\frac{q_n}{\epsilon_1} \frac{1}{\sqrt{({\bf r}_{e,\parallel}-{\bf r}_{h,\parallel})^2+
        [z_e-(-1)^n\, z_h-n\, L_z]^2}}
\end{equation}
\normalsize
\noindent where $q_n=(\frac{\epsilon_1-\epsilon_2}{\epsilon_1+\epsilon_2})^{|n|}$, $\epsilon_1$, $\epsilon_2$ are the nanoplatelet and surroundings dielectric constants, $L_z$ is the nanoplatelet height and ${\bf r}_{e,\parallel}$, ${\bf r}_{h,\parallel}$ the in-plane vector position of electron and hole. Please note that in the above equation (\ref{eqIII.22}) we disregard the image charges originated at the remote vertical nanoplatelet faces because their contribution is negligible and only account for those produced at the close horizontal ones located at a short $L_z/2$ distance. Then, integral (\ref{eqIII.21}) becomes,
\small
\begin{flalign} 
\label{eqIII.23}
& V_c = N^2 \; \iiiint\displaylimits_{-L/2}^{\;\;L/2} dx_e dx_h dy_e dy_h \cos^2 k x_e \cos^2 k x_h  \; * & \nonumber \\ 
&\hspace{0.5cm} * \; \cos^2 k y_e \cos^2 k y_h \; e^{-2 a \sqrt{(x_e-x_h)^2+(y_e-y_h)^2}}   * & \nonumber \\  
&\hspace{0.5cm} * \iint\displaylimits_{-L_z/2}^{\;\;\; L_z/2} dz_e dz_h \, H_c \, \cos^2 \bar{k} z_e \cos^2 \bar{k} z_h &
\end{flalign} 
\normalsize

\noindent First, we will turn the fourfold integral in $x_e, x_h, y_e, y_h$ into a twofold one by means a double analytical integration. To this end we start by defining $k x_e=\xi_e$, $k x_h=\xi_h$ so that, since $k=\pi/L$, the original $[-L/2,L/2]$ integration limits become $[-\pi/2,\pi/2]$. Next, we define $\xi=\xi_e-\xi_h$ and $\xi'= \xi_e+\xi_h$ so that,\cite{Romestain}
\small
\begin{flalign} 
\label{eqIII.24}
& \iint\displaylimits_{-L/2}^{\;\;L/2} dx_e dx_h  \cos^2 k x_e \cos^2 k x_h \, f(|x_e-x_h|) = & \nonumber \\ 
& \hspace{0.4cm} = \; \frac{1}{k^2}  \iint\displaylimits_{-\pi/2}^{\;\;\pi/2}  d\xi_e d\xi_h \cos^2  \xi_e \cos^2 \xi_h \, f(|\xi_e-\xi_h|) & \nonumber \\
& \hspace{0.4cm} = \;  \frac{1}{4 k^2}\int_0^{\pi} d\xi\,[(\pi-\xi) (2+\cos 2 \xi)+\frac{3}{2}\, \sin 2\xi] \, f(\xi)&
\end{flalign} 
\normalsize
\noindent For details on the last step in (\ref{eqIII.24}) see Appendix D. We do the same tranformation to the $y$ coordinates and refer either new coordinate to as $\xi_x$ and $\xi_y$. Then, with the notation $g(\xi)=(\pi-\xi) (2+\cos 2 \xi)+\frac{3}{2}\, \sin 2\xi$, the Coulomb integral (\ref{eqIII.23}) can be rewritten as:
\small
\begin{equation}
\label{eqIII.25}
V_c = \frac{N^2}{16} \frac{1}{k^4}   \iint\displaylimits_{0}^{\;\;\pi} g(\xi_x) \, g(\xi_y)  \exp[-\frac{2 a}{k} \sqrt{\xi_x^2+\xi_y^2}] \,*\, {\bf I}_2
\end{equation}
\normalsize
\noindent with
\small                                                                                                    
\begin{flalign} 
\label{eqIII.26}
&{\bf I}_2 = \iint\displaylimits_{-L_z/2}^{\;\;L_z/2}  dz_e dz_h   \, H_c(\xi_x,\xi_y,z_e,z_h) \,\cos^2 \bar{k} z_e \cos^2 \bar{k} z_h & \nonumber \\
& = \; \frac{1}{\bar{k}^2} \iint\displaylimits_{-\pi/2}^{\;\;\pi/2}  d\xi_{ze} d\xi_{zh} \, \, H_c(\xi_x,\xi_y,\xi_{ze},\xi_{zh}) \,
     \,\cos^2 \xi_{ze} \cos^2 \xi_{zh}&
\end{flalign} 
\normalsize
\noindent where now the n-th $H_c$ term reads:
\small
\begin{equation}
\label{eqIII.27}
\frac{q_n}{\epsilon_1} \, \frac{1}{\sqrt{(1/k^2) (\xi_x^2+\xi_y^2)+(1/\bar{k}^2) \, [\xi_{ze}-(-1)^n\, \xi_{zh} - n \pi]^2}}.
\end{equation}
\normalsize
\noindent The integral ${\bf I}_2$  cannot be calculated analytically. We then calculate ${\bf I}_2$ semi-analytically. To this end we consider that
\small 
\begin{equation}
\label{eqIII.28}
\int_a^b f(x) g(x) dx =\sum_{i=1}^{N-1} g(a+(i-1/2) \Delta) \; \int_{a+(i-1)\Delta}^{a+i\,\Delta} f(x) dx
\end{equation}
\normalsize
\noindent where $g(x)$ is a smooth function while $f(x)$ may have regions with sharp gradients. This strategy  is in the core of the envelope function $k\cdot p$ model widely employed to describe semiconductor heterostructures and quantum dots.\cite{Bastard}

\noindent In our case the smooth function is the product of cosines. Then, should we withdraw this product the resulting integral has primitive:
\small
\begin{flalign} 
\label{eqIII.29}
& P(\xi_{ze},\xi_{zh})=\iint \frac{ d\xi_{ze} d\xi_{zh}}{\sqrt{a^2+[\xi_{ze}-(-1)^n\, \xi_{zh} - n \pi]^2}} \; =& \nonumber \\
&\hspace{0.1cm} = \; (-1)^n \left\{ \sqrt{a^2+(n \pi-\xi_{ze}+(-1)^n\, \xi_{zh})^2} \right. +& \nonumber \\
&\hspace{0.5cm} + \; (-1)^n \, \xi_{zh} \, \log \left[ - n\pi+\xi_{ze}-(-1)^n\, \xi_{zh}+ \right. & \nonumber \\
&\hspace{2cm} \left. + \;\sqrt{a^2+(n \pi-\xi_{ze}+(-1)^n\, \xi_{zh})^2} \right] & \nonumber \\
&\hspace{0.5cm}  + \; (\xi_{ze}-n \pi) \,  \log \left[ n\pi-\xi_{ze}+(-1)^n\, \xi_{zh} + \right. & \nonumber \\
&\hspace{2cm} \left. \left. +\; \sqrt{a^2+(n \pi-\xi_{ze}+(-1)^n\, \xi_{zh})^2}\right] \right\} &
\end{flalign} 
\normalsize                                                                 
\noindent Then, we divide symmetrically the interval $[-\pi/2,\pi/2]$ in a pair number $N$ of subintervals and in each subinterval we replace the smooth function $g(x)$ by its value at the center of it $g(a+(i-1/2) \Delta)$ and write the integral with limits $[a,b]$ as a sum of $N$ analytical functions. In our case, after labeling as $\xi_{ze}^{(i)}, \xi_{ze}^{(f)},\xi_{zh}^{(i)},\xi_{zh}^{(f)}$ the limits of the subinterval, the result of the integral in it is:
\small
\begin{flalign} 
\label{eqIII.30}
& \cos^2 [\frac{1}{2}(\xi_{ze}^{(f)}+\xi_{ze}^{(i)})] \cos^2 [\frac{1}{2}(\xi_{zh}^{(f)}+\xi_{zh}^{(i)})] \; * & \nonumber \\ 
& \hspace{1cm} *\; \left[P(\xi_{ze}^{(f)},\xi_{zh}^{(f)})+P(\xi_{ze}^{(i)},\xi_{zh}^{(i)}) \right. - & \nonumber \\ 
& \hspace{2cm} \left. - \, P(\xi_{ze}^{(f)},\xi_{zh}^{(i)})-P(\xi_{ze}^{(i)},\xi_{zh}^{(f)}) \right]. &
\end{flalign} 
\normalsize 
\noindent The result of integration in $z$ is then a sum of terms only dependent on $\xi_x, \xi_y$ that we refer to as $Z(\xi_x, \xi_y)$. The usefulness of the application of (\ref{eqIII.28}) in our case is that $Z(\xi_x, \xi_y)$ and then the Coulomb integral is highly convergent with the number $N$ of subintervals.\cite{Movilla}\\

\noindent From the above algebra, the sixfold Coulomb integral turns into the following numerical twofold one:
\small
\begin{equation}
\label{eqIII.31}
V_c = \frac{N^2}{16} \frac{1}{k^4} \iint\displaylimits_{0}^{\;\;\;\;\pi} d\xi_x d\xi_y \,g(\xi_x) g(\xi_y) Z(\xi_x,\xi_y) e^{-\frac{2 a}{k}\sqrt{\xi_x^2+\xi_y^2}}
\end{equation}
\normalsize 
\noindent In the case of a rectangular nanoplatelet we proceed in a similar way, just taking into account that now instead of a unique $k=\pi/L$ we have $k_x=\pi/L_x$ and $k_y=\pi/L_y$.

\subsection{Long nanorods: a quasi-1D system}
\label{subsec:3.2}
The exciton variational wave function is also chosen to be a product of the electron and hole lowest-energy subband states and a Slater correlation factor.
\small
\begin{flalign} 
\label{eqIV.1}
& \Psi(r_e, r_h)= N \cos k x_e \cos k x_h \cos \bar{k} y_e \cos \bar{k} y_h \;* &\nonumber\\
&\hspace{3cm} \cos \bar{k} z_e \cos \bar{k} z_h  \; e^{-a |x_e-x_h|}&
\end{flalign} 
\normalsize 
\noindent Where $N$ is the normalization factor, $k= \pi/L_x$ and $\bar{k}= \pi/L_y = \pi/L_z$. This wave-function also gives the correct results for the ground state of the exciton in the limits of extremely high (small $L_x$) and negligible (large $L_x$) confinement. 

\noindent As above, the use of a Slater correlation factor has the advantage of having a simple additive closed-form for the kinetic energy (see Appendix A for details):
\small
\begin{equation}
\label{eqIV.2}
\langle \Psi(r_e,r_h)| \hat T|  \Psi(r_e,r_h)\rangle = \frac{\bar{k}^2}{2 \mu_z}+\frac{\bar{k}^2}{2\mu_{\parallel}}+\frac{k^2}{2\mu_{\parallel}}+\frac{a^2}{2\mu_{\parallel}}
\end{equation}
\normalsize 
\noindent with $k= \pi/L_x $ and $\bar{k} = \pi/L_y = \pi/L_z$ and $a$ the variational parameter to be optimized.

\subsubsection{The single-particle density}
\label{subsubsec:3.2.1}

In this section we also employ the labels 1 and 2 to refer to either particle and $L$ to refer to $L_y=L_z$. In Appendix C we enclose the value of the integrals employed to derive the kinetic energy, density and norm of the above wave-function. The single-particle density of particle "1" reads,
\small
\begin{eqnarray}
\label{eqIV.3}
\rho(r_1) &=& \int |\Psi(r_1,r_2)|^2 d^3r_2 = N^2 \; (\frac{L}{2})^2 \cos^2 \bar{k} y_1 \cos^2 \bar{k} z_1 \; *\nonumber \\
           &*&   \int_{-L_x/2}^{L_x/2} \cos^2k x_1 \cos^2 k x_2 e^{-2 a \,|x_2-x_1|} \; dx_2 
\end{eqnarray}
\normalsize
\noindent Since (see Appendix C)
\small
\begin{flalign} 
\label{eqIV.4}
& \int_{-L_x/2}^{L_x/2} \cos^2 k x_2 \; e^{-2 a \,|x_2-x_1|} \; dx_2 \; = & \nonumber \\
&\hspace{3cm} =\; \frac{1}{2 a}+\frac{a}{2\, (a^2+k^2)} \cos 2 k x_1 &
\end{flalign} 
\normalsize
\noindent the single-particle density results:
\small
\begin{flalign} 
\label{eqIV.5}
& \rho(r_1) = N^2 \;  (\frac{L}{2})^2 \; \left[ \frac{1}{2 a}+\frac{a}{2\,(a^2+k^2)} \cos 2 k x_1 \right] \; * & \nonumber\\
&\hspace{3cm} *\; \cos^2 \bar{k} y_1 \; \cos^2 \bar{k} z_1 \; \cos^2 k x_1 &
\end{flalign} 
\normalsize
\noindent Finally, by integrating $\rho(r_1)$, we get the norm:
\small
\begin{equation}
\label{eqIV.6}
N^2 = (\frac{2}{L})^4\, (\frac{2}{L_x})\, \frac{8 a \, (a^2 + k^2)}{6 a^2+4 k^2}
\end{equation}
\normalsize

\subsubsection{The bare-Coulomb integral}
\label{subsubsec:3.2.2}

We show here that the sixfold integral, 
\small
\begin{equation}
\label{eqIV.7}
V_c=\int \frac{|\Psi({\bf r}_e,{\bf r}_h)|^2}{\epsilon \, |{\bf r}_e-{\bf r}_h|} \; d^3{\bf r}_e d^3 {\bf r}_h
\end{equation}
\normalsize
\noindent where $\Psi({\bf r}_e,{\bf r}_h)$ is given in eq. (\ref{eqIV.1}), can be reduced up to a one-coordinate numerical integration. To this end we first turn the sixfold integral in $x_e, x_h, y_e, y_h, z_e, z_h$ into a threefold one by using the set of variables $\xi_x$, $\xi'_x$,$\xi_y$, $\xi'_y$,$\xi_z$, $\xi'_z$, with $\xi_x= k (x_e-x_h)$, $\xi'_x= k (x_e+x_h)$ $\dots$  $\xi'_z= {\bar k} (z_e+z_h)$, and carry out analytical integrations over $\xi'_i \,$\cite{Romestain} yielding (see Appendix D for details),
\small
\begin{flalign} 
\label{eqIV.8}
& V_c = \frac{N^2}{2^6} \frac{1}{k^2\, \bar{k}^4} \frac{1}{\epsilon} \iiint\displaylimits_{0}^{\hspace{0.3cm} \pi} d\xi_x d\xi_y d\xi_z \,g(\xi_x) g(\xi_y) g(\xi_z) \; * & \nonumber \\
&\hspace{2.5cm} *\; e^{-\frac{2 a}{k}\,\xi_x} \; \frac{\bar{k}}{\sqrt{(\bar{k}/k)^2\, \xi_x^2+\xi_y^2+\xi_z^2}}
\end{flalign} 
\normalsize
\noindent with $g(\xi_i)=(\pi-\xi_i) (2+\cos 2 \xi_i)+\frac{3}{2}\, \sin 2\xi_i$.\\

\noindent  Now, we may numerically integrate this threefold integral. Alternatively, we can consider the primitive $P(\xi_y,\xi_z)$ of the next integral,
\small
\begin{flalign} 
\label{eqIV.9}
& P(\xi_y,\xi_z)=\iint\frac{d\xi_y d\xi_z}{\sqrt{a^2+\xi_y^2+\xi_z^2}} =  -\xi_y + a\, \arctan\frac{\xi_y}{a} \; - &\nonumber\\
& - a \,\arctan\frac{\xi_y \xi_z}{a\,\sqrt{a^2+\xi_y^2+\xi_z^2}} + \xi_z \log(\xi_y+\sqrt{a^2+\xi_y^2+\xi_z^2}\;)\;+&\nonumber \\
&+ \xi_y \log(\xi_z+\sqrt{a^2+\xi_y^2+\xi_z^2}\;)&
\end{flalign} 
\normalsize
\noindent where $a=\frac{\bar{k}}{k}\, \xi_x$, that can also be written as,
\small
\begin{flalign}
\label{eqIV.10}
& P(\xi_y,\xi_z)=\iint\frac{d\xi_z d\xi_y}{\sqrt{a^2+\xi_y^2+\xi_z^2}} = -\xi_z + a\, \arctan\frac{\xi_z}{a} \; - &\nonumber\\
&- a \,\arctan\frac{\xi_y \xi_z}{a\,\sqrt{a^2+\xi_y^2+\xi_z^2}} + \xi_z \log(\xi_y+\sqrt{a^2+\xi_y^2+\xi_z^2}\;)\;+&\nonumber \\
&+ \xi_y \log(\xi_z+\sqrt{a^2+\xi_y^2+\xi_z^2}\;).&
\end{flalign}
\normalsize
\noindent Please note that for any square interval of limits $\{[\xi_y^{(i)},\xi_y^{(f)}]$, $[\xi_z^{(i)},\xi_z^{(f)}]\}$, the definite integral
$[P(\xi_y^{(f)},\xi_z^{(f)})+ P(\xi_y^{(i)},\xi_z^{(i)})-P(\xi_y^{(f)},\xi_z^{(i)})-P(\xi_y^{(i)},\xi_z^{(f)})]$ is the same, irrespectively of using $P(\xi_y,\xi_z)$ according to eq. (\ref{eqIV.9}) or eq. (\ref{eqIV.10}).\\

\noindent This result allows to perform the integral,
\small
\begin{equation}
\label{eqIV.11}
Z(\xi_x)= \iint\displaylimits_{0}^{\;\;\;\;\;\;\pi} d\xi_y d\xi_z \, g(\xi_y)\, g(\xi_z)\, \frac{\bar{k}}{\sqrt{(\bar{k}/k)^2\, \xi_x^2+\xi_y^2+\xi_z^2}}
\end{equation}
\normalsize
\noindent as a sum of terms like:
\small
\begin{flalign}
\label{eqIV.12}
& g[\frac{1}{2}(\xi_z^{(i)}+\xi_z^{(f)})]\, g[\frac{1}{2}(\xi_y^{(i)}+\xi_y^{(f)})]\;[P(\xi_y^{(f)},\xi_z^{(f)})+\nonumber\\
&\hspace{0.75cm}+\;  P(\xi_y^{(i)},\xi_z^{(i)}) -P(\xi_y^{(f)},\xi_z^{(i)})-P(\xi_y^{(i)},\xi_z^{(f)})].&
\end{flalign}
\normalsize
\noindent Finally, we numerically obtain the bare Coulomb term by carrying out the one-coordinate integral:
\small
\begin{equation}
\label{eqIV.13}
\int_0^{\pi} d\xi_x \, g(\xi_x)\, Z(\xi_x)\; e^{-\frac{2 a}{k}\,\xi_x}. 
\end{equation}
\normalsize

\subsection{Cubic quantum dots: a 0D system}
\label{subsec:3.3}

As in the above sections, the exciton variational wave function is chosen to be a product of the electron and hole lowest-energy subband states and a Slater correlation factor. 
\small
\begin{flalign}
\label{eqV.1}
& \Psi(r_e, r_h)= N \cos k x_e \cos k x_h \cos k y_e \cos k y_h \cos k z_e \;*\nonumber\\
&\hspace{1cm} *\, \cos k z_h \; e^{-a \sqrt{(x_e-x_h)^2+(y_e-y_h)^2++(z_e-z_h)^2} }&
\end{flalign}
\normalsize
\noindent This variational wave-function has been previously employed by Romestain and Fishman.\cite{Romestain}  We follow similar techniques as those employed in the previous sections to derive closed form of the kinetic energy, single-particle density and the normalization factor. We present next the model for cubic QDs. The extension to cuboid QDs is straightforward with the help of integrals in Appendix C. 

\subsubsection{The single-particle density}
\label{subsubsec:3.3.1}

The single-particle density of particle "1" reads,

\small
\begin{flalign}
\label{eqVa.2}
& \rho(r_1) = \int |\Psi(r_1,r_2)|^2 d^3r_2 = N^2 \, \cos^2 k x_1 \cos^2 k y_1 \cos^2 k z_1 * &\nonumber \\
&\hspace{1cm} *\; \iiint\displaylimits_{-L/2}^{\;\;\;L/2} d x_2\,d y_2\,d z_2\, \cos^2 k x_2 \cos^2 k y_2 \cos^2 k z_2 \;* &\nonumber \\
&\hspace{2cm} *\; e^{-2 a \,\sqrt{(x_2-x_1)^2+(y_2-y_1)^2+(z_2-z_1)^2}} &
\end{flalign}
\normalsize

\noindent We obtain in Appendix B the following closed-form for the single-particle density: 
\small
\begin{flalign}
\label{eqVa.9}
& \rho(r_1) = N^2 \, \frac{\pi}{8} \; \cos^2 k x_1 \; \cos^2 k y_1 \; \cos^2 k z_1  \;* & \nonumber \\
&\hspace{0.5cm} * \;  \left\{ \frac{1}{a^3}+\frac{a}{(a^2+k^2)^2} \left(\cos 2 k x_1  + \cos 2 k y_1  +\cos 2 k z_1 \right) 
              \right. & \nonumber   \\
&\hspace{1cm}  + \frac{a}{(a^2+2k^2)^2} \left[ \cos 2 k x_1 \, \cos 2 k y_1  \; + \right. & \nonumber   \\
&\hspace{2.5cm} \left. +\cos 2 k x_1 \, \cos 2 k z_1 + \, \cos 2 k y_1 \, \cos 2 k z_1 \right] \nonumber   \\
&\hspace{1cm}  \left. + \frac{a}{(a^2+3k^2)^2} \; \cos 2 k x_1 \, \cos 2 k y_1  \, \cos 2 k z_1 \right\}&
\end{flalign}
\normalsize  
\noindent The integration of $\rho(r_1)$ from $-L/2$ up to $L/2$ in all three coordinates should yield the unity. Then, since $k=\pi/L$, we get the norm:
\small 
\begin{flalign}
\label{eqV.10} 
& N^2 = \frac{512 \, a^3}{\pi \, L^3}\; \left(8+\frac{12}{(1+(\frac{k}{a})^2)^2} \, + \right. &\nonumber\\
&\hspace{2cm}\left. +\frac{6}{(1+2\,(\frac{k}{a})^2)^2} + \frac{1}{(1+3\,(\frac{k}{a})^2)^2}\right)^{-1}&
 \end{flalign}
\normalsize  
\noindent In order to extend the previous model to cuboid QDs with edges of different length we would just make use of the last two eqs. in Appendix C.2. All the same, as pointed out above, if the lengths of the cuboid QD are not similar, the use of a single variational parameter model may not be enough to reach the same accuracy than that reached for cubic QDs. In order to keep the accuracy, the correlation factor should be supplied with additional variational parameters. Namely, $f(r_{eh})=\exp[-a\, \sqrt{(x_e-x_h)^2+ b\, (y_e-y_h)^2+ c\, (z_e-z_h)^2}]$. However, the three-parameters model has not the simplicity of the single-parameter one: no closed formulas for kinetic energy, density
and normalization factors can be found in this case and, therefore, the calculation becomes much heavier, unable for extensive calculations.

\section{Summary}
\label{section:summary}
We derive closed-form formulas for the normalization factor, single-particle density and expectation value of the kinetic energy for simple correlated exciton wave-functions chosen to be a product of the electron and hole lowest-energy subband states times a Slater correlation factor suited to describe 0D, quasi-1D and quasi-2D systems. We also provide fast integration procedures for the Coulomb integral in typical quasi-2D nanoplate\-lets including polarization of the Coulomb interaction, and for the bare-Coulomb integral in quasi-1D long nanorods. For nano\-platelets, the original sixfold integral is reduced to a twofold numerical integral, while a one-coordinate numerical integration is required for quasi-1D systems. This theoretical baggage should enable accurate yet reliable simulations of electronic structure in strongly correlated exciton systems of current interest.

\begin{acknowledgements}
I thank J. Karwowski for most useful discussions and F. Rajadell for a careful revision of the derivations and formulas.
\end{acknowledgements}

%%%%%%%%%%%%%%%%%%%%%%%%%%%%%%%%%%%%%%%%%%%%%%%%%%%%%%%%%%%%%%%%%%%%%%%%%%%%%%%%%%%
\appendix
\section{Deriving the expectation value of the kinetic energy for the different models}
\label{section:5}
\subsection{Cuboid QD and nanoplatelet}
\label{subsection:5.1}
\normalsize 
The normalized wave-function for these models is $\Psi= \Phi({\bf r}_e,{\bf r}_h) \chi(r_{eh})$ with  
$\Phi({\bf r}_e,{\bf r}_h)=\phi_e({\bf r}_e) \, \phi_h({\bf r}_h)$ and $\chi(r_{eh})$ $= \exp[-a \, r_{eh}]$. 
In the 0D model $r_{eh}$ is three-dimensional i.e. $r_{eh}=\sqrt{(x_e-x_h)^2+(y_e-y_h)^2+(z_e-z_h)^2}$ 
while the the quasi-2D model it is two-dimensional i.e.,
$r_{eh}=\sqrt{(x_e-x_h)^2+(y_e-y_h)^2}$.\\
 
\noindent The most general kinetic energy operator reads,
\small 
\begin{equation}
\label{Ap1.1} 
 \hat T = \sum_{i=e,h} \; \sum_{a=x,y,z} -\frac{1}{2 m_{ia}} \frac{\partial^2}{\partial a_{i}^2} 
\end{equation}
\normalsize                                                                                                           
\noindent In order to simplify the notation we define $D=\frac{\partial}{\partial  a_{i}}$ and $w=\langle \Psi| D^2|\Psi \rangle$. We have,
\small 
\begin{equation}
\label{Ap1.2} 
D^2 (F \,G) = (D^2 \, F) \, G+2\, (D F)(D G)+F\,(D^2G)
\end{equation}
\begin{equation}
\label{Ap1.3} 
w=\langle \Phi \; \chi^2|D^2 \Phi\rangle +\langle \Phi^2 \, \chi|D^2 \chi \rangle + 2 A
\end{equation}
\normalsize 
\noindent with,
\small  
\begin{equation}
\label{Ap1.4} 
A=\langle \Phi \, \chi|(D \Phi)(D \chi)\rangle = \frac{1}{4} \, \langle (D \Phi^2)(D \chi^2) \rangle.
\end{equation} 
\normalsize  
\noindent Since  $\Phi=0$ at the integration limit, after integration by parts,
\small 
\begin{equation}
\label{Ap1.5} 
A=-\frac{1}{4} \; \langle \Phi^2|(D ^2 \chi^2)\rangle.
\end{equation} 
\normalsize  
\noindent Finally,
\small  
\begin{equation}
\label{Ap1.6} 
W= \langle \Psi|\hat T|\Psi\rangle = \langle \Phi \, \chi^2|\hat T|\Phi\rangle + \langle \Phi^2 | \chi \, (\hat T \chi) -\frac{1}{2} (\hat T \chi^2)\rangle.
\end{equation} 
\normalsize   
\noindent We have that $\hat T \Phi = E \Phi$.  Since $\frac{1}{m_e}+\frac{1}{m_h}=\frac{1}{\mu}$, then 
\small  
\begin{equation}
\label{Ap1.7} 
E=\frac{k_x^2}{2 \mu_x}+\frac{k_y^2}{2 \mu_y}+\frac{k_z^2}{2 \mu_z}
\end{equation} 
\normalsize 
\noindent In the case of a cubic 0D QD with isotropic mass $E=\frac{3 k^2}{2 \mu}$ while $E=\frac{1}{2 \mu}\, (k_x^2+k_y^2+k_z^2)$ for a cuboid QD. For a platelet with anisotropic mass $E=\frac{k^2}{\mu_{\parallel}}+\frac{k_z}{2 \mu_z}$.\\

\noindent We calculate next the second integral in eq. (\ref{Ap1.6}) for the case of an isotropic QD. To this end we use center-of-mass and relative motion coordinates: ${\bf R}=(m_e {\bf r}_e+m_h {\bf r}_h)/M$, ${\bf r}={\bf r}_e-{\bf r}_h$. We have,
\small 
\begin{equation}
\label{Ap1.8}                                                         
\hat T = \frac{P^2}{2 M}+\frac{p^2}{2 \mu}= \hat T_R + \hat T_r 
\end{equation} 
\normalsize 
\noindent Since $\chi = \exp[-a r]$, $\hat T \, \chi =\hat T_r \, \chi=-\frac{1}{2}\frac{1}{r^2}\frac{d}{dr}(r^2\frac{d}{dr}) \; \chi$. Then,
\small
\begin{equation}
\label{Ap1.9}                                                         
\chi\, \hat T \, \chi - \frac{1}{2}\, \hat T \,\chi^2 = \chi\, \hat T_r \, \chi - \frac{1}{2}\, \hat T_r \,\chi^2 = \frac{a^2}{2\mu} \; \chi^2
\end{equation}  
\normalsize 
 
\noindent and
\small
\begin{equation}
\label{Ap1.10}                                                         
W=\frac{3 \, k^2}{2 \,\mu}+\frac{a^2}{2 \,\mu}
\end{equation}  
\normalsize 
 
\noindent In the case of a nanoplatelet the correlation factor is two-dimensional ($r_{eh}=\sqrt{(x_e-x_h)^2+(y_e-y_h)^2}\;$). We follow now similar steps as above but instead of spherical we use polar $(r,\theta, z)$ coordinates. In this case,
\small 
\begin{equation}
\label{Ap1.11}                                                         
\hat T_r = -\frac{1}{2 \mu_{\parallel}}\; (\frac{\partial^2}{\partial r^2}+\frac{1}{r} \frac{\partial}{\partial r}) 
\end{equation} 
\normalsize 

\noindent  and 
\small
\begin{equation}
\label{Ap1.12}                                                         
\chi\, \hat T \, \chi - \frac{1}{2}\, \hat T \,\chi^2 = \chi\, \hat T_r \, \chi - \frac{1}{2}\, \hat T_r \,\chi^2 = \frac{a^2}{2\mu_{\parallel}} \; \chi^2
\end{equation}   
\normalsize 
 
\noindent  So that finally, 
\small 
\begin{equation}
\label{Ap1.13}                                                         
W=\frac{k^2}{\mu_{\parallel}}+\frac{k_z^2}{2 \,\mu_z}+\frac{a^2}{2 \,\mu_{\parallel}}.
\end{equation}   
\normalsize 

\subsection{Long nanorod}
\label{subsection:5.2}

The wave-function in this case, eq. (\ref{eqIV.1}), is: 
\small
\begin{flalign} 
& \Psi(r_e, r_h)= N \cos k x_e \cos k x_h \cos \bar{k} y_e \cos \bar{k} y_h \;* &\nonumber\\
&\hspace{3cm} *\; \cos \bar{k} z_e \cos \bar{k} z_h  \; e^{-a |x_e-x_h|}&\nonumber
\end{flalign} 
\normalsize 
\noindent It is immediate to find that, 
\small 
\begin{equation}
\label{Ap1.14}                                                         
\frac{\langle \Psi|\hat T_y|\Psi \rangle}{\langle \Psi|\Psi \rangle} = \frac{\bar{k}^2}{2 \mu_{\parallel}}
\end{equation}  
\normalsize  
\noindent with $\hat T_y=-\frac{1}{2 m_{e,\parallel}} \,\frac{\partial^2}{\partial y_e^2}\, -\frac{1}{2 m_{h,\parallel}}\,\frac{\partial^2}{\partial y_h^2}$ and $\frac{1}{m_e}+\frac{1}{m_h}=\frac{1}{\mu}$.\\

\noindent Similarly we find that,
\small
\begin{equation}
\label{Ap1.15}                                                         
\frac{\langle \Psi|\hat T_z|\Psi \rangle}{\langle \Psi|\Psi \rangle} = \frac{\bar{k}^2}{2 \mu_z}
\end{equation}  
\normalsize 
 
\noindent For the long edge, with $\hat T_x=-\frac{1}{2 m_{e,\parallel}} \,\frac{\partial^2}{\partial x_e^2}\, -\frac{1}{2 m_{h,\parallel}}\,\frac{\partial^2}{\partial x_h^2}$, 
$\langle \Psi|\Psi \rangle=1$, we have,
\small 
\begin{flalign} 
\label{Ap1.16}                                                         
& \frac{\langle \Psi|\hat T_x|\Psi \rangle}{\langle \Psi|\Psi \rangle} = N^2 \, \int_{-\frac{L}{2}}^{\frac{L}{2}} d y_e \, \cos^2 \bar{k} y_e \, \int_{-\frac{L}{2}}^{\frac{L}{2}} d y_h  \, \cos^2 \bar{k} y_h \,*&\nonumber\\ 
&\hspace{0.5cm} *\; \int_{-\frac{L}{2}}^{\frac{L}{2}} d z_e  \, \cos^2 \bar{k} z_e \, \int_{-\frac{L}{2}}^{\frac{L}{2}} d z_h  \, \cos^2 \bar{k} z_h \; * &\nonumber \\
&\hspace{0.5cm} * \iint\displaylimits_{-\frac{L_x}{2}}^{\;\;\;\;\frac{L_x}{2}}  d x_e d x_h  \cos k x_e \, \cos k x_h \; e^{-a\,|x_e-x_h|} \;* &\nonumber \\
&\hspace{2.5cm} \hat T_x \, \cos k x_e \, \cos k x_h \exp[-a\,|x_e-x_h|] &\nonumber \\
			 &= N^2 \, (\frac{L}{2})^4 \; ({\bf I}_1+{\bf I}_2)&
\end{flalign} 
\normalsize 
\noindent Since 
\small
\begin{flalign} 
\label{Ap1.17}                                                         
&\frac{d^2}{d x_e^2} \left( \cos k x_e \,e^{-a \,|x_h-x_e|} \right) =  e^{-a \, |x_h-x_e|}\, *& \nonumber \\
&\hspace{1cm}*\, [(a^2-k^2) \, \cos k x_e + 2 a k \, \frac{|x_h-x_e|}{x_h-x_e} \,\sin k x_e  & \nonumber \\
&\hspace{3.5cm}  - 2 a \, \cos k x_e \, \delta(x_h-x_e)]&
\end{flalign}
\normalsize  
\noindent The integral ${\bf I}_1$ becomes split into a sum of three integrals, ${\bf I}_1= {\bf I}_{11}+{\bf I}_{12}+{\bf I}_{13}$, where (see Appendix C.1 for the employed definite integrals), 
\small 
\begin{eqnarray}
\label{Ap1.18}                                                         
{\bf I}_{11}&=&-\frac{a^2-k^2}{2 m_e} \, \int_{-\frac{L_x}{2}}^{\frac{L_x}{2}} d x_e \, \cos^2 k_e \; * \nonumber\\
            &&\hspace{1.5cm} *\; \int_{-\infty}^{\infty} d x_h \, \cos k x_h \exp[-2 a \,|x_h-x_e|] \nonumber\\
            &=&-\frac{a^2-k^2}{2 m_e} \, \int_{-\frac{L_x}{2}}^{\frac{L_x}{2}} d x_e \, \cos^2 k_e \; \left(\frac{1}{2 a}+\frac{a \, \cos 2 k x_e}{2 \,(a^2+k^2)} \right) \nonumber\\
            &=&-\frac{a^2-k^2}{2 m_e} \, \left\{ \frac{1}{2a}\,\int_{-\frac{L_x}{2}}^{\frac{L_x}{2}} d x_e \, \cos^2 k_e \;\right. +\nonumber\\
			&&\hspace{1cm} \left. +\,\frac{a}{2\, (a^2+k^2)} \,\int_{-\frac{L_x}{2}}^{\frac{L_x}{2}} d x_e \, \cos^2 k x_e \, \cos 2 k x_e \right\} \nonumber\\
			&=&	-\frac{a^2-k^2}{2 m_e} \, \frac{L_x}{2} \; \left( \frac{1}{2a}+\frac{a/4}{a^2+k^2}\right)
\end{eqnarray}
\normalsize 
 
\small
\begin{eqnarray}
\label{Ap1.19}                                                         
{\bf I}_{12}&=& - \frac{a \, k}{m_e}\;\int_{-\frac{L_x}{2}}^{\frac{L_x}{2}} d x_h \, \cos^2 k x_h \, *\nonumber\\
             &*& \int_{-\infty}^{\infty} d x_e\, \cos k x_e \, \sin k x_e\,
                \frac{|x_h-x_e|}{x_e-x_h}\; e^{-2 a |x_h-x_e|}
\end{eqnarray}
\normalsize 
\noindent Since (see Appendix C.1),  
\small 
\begin{flalign} 
\label{Ap1.20}                                                         
&\int_{-\infty}^{\infty} d \bar{x}_e\, \sin 2 k (\bar{x}_e+x_h) \frac{|\bar{x}_e|}{\bar{x}_e}\; \exp[-2 a |\bar{x}_e|] = &\nonumber\\
&\hspace{3.5cm} = \;\; \frac{k}{a^2+k^2}\;\cos 2 k x_h&
\end{flalign}
\normalsize 
\noindent then,
\small 
\begin{eqnarray}
\label{Ap1.21}                                                         
{\bf I}_{12}&=& - \frac{a \, k}{m_e} \; \frac{1}{2} \, \frac{k}{a^2+k^2} \; \int_{-\frac{L_x}{2}}^{\frac{L_x}{2}} d x_h \, \cos^2 k x_h \, \cos 2 k x_h \nonumber \\
			&=& - \frac{a \, k^2}{2 m_e}\, \frac{1}{a^2+k^2} \, \frac{L_x}{4}.
\end{eqnarray}
\normalsize 
 
\noindent Now we integrate  ${\bf I}_{13}$:
\small  
\begin{eqnarray}
\label{Ap1.22}                                                         
{\bf I}_{13}&=& \frac{a}{m_e} \, \int\displaylimits_{-\frac{L_x}{2}}^{\;\;\;\frac{L_x}{2}} d x_e \, d x_h \, \; \delta(x_h-x_e)\;  \cos^2 k x_e \;*\nonumber \\
&& \hspace{3cm} * \;\cos^2 k x_h \; e^{-2 a |x_h-x_e|}  \nonumber \\
		    &=& \frac{a}{m_e} \; \int_{-\frac{L_x}{2}}^{\frac{L_x}{2}} d x_e \, \cos^2 k x_e \;*\nonumber \\  
			&& * \; \int_{-\infty}^{\infty} d x_h \, \cos^2 k x_h 
			     \; \exp[-2 a |x_h-x_e|] \; \delta(x_h-x_e) \nonumber \\
		    &=& \frac{a}{m_e} \; \int_{-\frac{L_x}{2}}^{\frac{L_x}{2}} d x_e \, \cos^2 k x_e  \, \cos^2 k x_e   \nonumber \\
			&=& \frac{3 a L_x}{8 m_e}.
\end{eqnarray}
\normalsize  
\noindent From eqs. (\ref{Ap1.18}), (\ref{Ap1.21}) and (\ref{Ap1.22}), we obtain ${\bf I}_1$:
\small 
\begin{equation}
\label{Ap1.23} 
 {\bf I}_1= \frac{L_x}{4 \,a} \left( \frac{k^2}{2 m_e}+\frac{3}{2} \frac{a^2}{2 m_e}\right)
\end{equation}
\normalsize 
 \noindent The integral ${\bf I}_2$ in eq. (\ref{Ap1.16}) is like ${\bf I}_1$ replacing $e$ by $h$. From ${\bf I}_1$ and ${\bf I}_2$, with $\frac{1}{m_e}+\frac{1}{m_h}=\frac{1}{\mu}$, we obtain
 ${\bf I}_1+{\bf I}_2= \frac{L_x}{4 a} \left( \frac{k^2}{2 \mu_{\parallel}}+\frac{3}{2} \frac{a^2}{2 \mu_{\parallel}}\right)$, so that:
\small 
\begin{eqnarray}
\label{Ap1.24}                                                         
\frac{\langle \Psi|\hat T_x|\Psi \rangle}{\langle \Psi|\Psi \rangle} &=& N^2 \, (\frac{L}{2})^4 \; ({\bf I}_1+{\bf I}_2)\nonumber  \\
                   &=& \frac{1}{2 \mu_{\parallel}}\; (a^2+k^2)
\end{eqnarray}
\normalsize 
\noindent Then,
\small
 \begin{eqnarray}
\label{Ap1.25}                                                         
\frac{\langle \Psi|\hat T|\Psi \rangle}{\langle \Psi|\Psi \rangle} = \frac{\bar{k}^2}{2 \mu_{\parallel}}+\frac{\bar{k}^2}{2 \mu_z} + \frac{k^2}{2 \mu_{\parallel}} + \frac{a^2}{2 \mu_{\parallel}}.
\end{eqnarray}
\normalsize 

%............................................................................................................................%

\section{The single-particle density}
\label{sec:6}
\subsection{The single-particle density of quasi-2D systems}
\label{subsec:6.1}
We employ the labels 1 and 2 to refer to either particle and $L$ to refer to $L_x=L_y$. The single-particle density of particle "1" reads,

\small
\begin{flalign} 
\label{eqIII.4}
& \rho(r_1) = \int |\Psi(r_1,r_2)|^2 d^3r_2 & \nonumber \\
& = N^2 \; \frac{L_z}{2} \cos^2 \bar{k} z_1 \cos^2 k x_1 \cos^2 k y_1 * & \nonumber \\
           & * \iint\displaylimits_{-L/2}^{\;\;L/2} \cos^2 k x_2 \cos^2 k y_2 \; e^{-2 a \,\sqrt{(x_2-x_1)^2+(y_2-y_1)^2}} d x_2 d y_2 & \nonumber \\
		   & =  N^2 \; \frac{L_z}{2} \cos^2 \bar{k} z_1 \cos^2 k x_1 \cos^2 k y_1 * \bf I
\end{flalign}
\normalsize

\noindent The identity $2 \cos^2 A = 1 + \cos 2 A$ removes the squares in the cosines of integral $\bf I$. Next, we carry out the following change of variables:
$x_2-x_1=\bar{x}_2$, $y_2-y_1=\bar{y}_2$, so that the pre-exponential factor in $\bf I$ becomes:

\small
\begin{flalign} 
\label{eqIII.5}
& \frac{1}{4}\left[1+ \cos(2 k \bar{x}_2+2 k x_1)+ \cos(2 k \bar{y}_2+2 k y_1) + \right. & \nonumber \\
& \hspace{1.5cm} \left. + \cos(2 k \bar{x}_2+2 k x_1)  \cos(2 k \bar{y}_2+2 k y_1)\right]
\end{flalign} 
\normalsize

\noindent Then the integral $\bf I$ can then be split as a sum of four integrals ${\bf I}_i$, $i=1,2,3,4$. Let's consider them separately.
\small
\begin{flalign}
\label{eqIII.6}
& {\bf I}_1=\frac{1}{4} \iint\displaylimits_{-L/2}^{\;\;\;L/2} e^{-2 a \,\sqrt{ \bar{x}_2^2+\bar{y}_2^2} \; } d \bar{x}_2 d \bar{y}_2 \approx &\nonumber \\
&\hspace{0.5cm} \approx  \frac{1}{4} \int_{0}^{\infty} \int_{0}^{2\pi}  e^{-2 a \bar{r}} \, \bar{r}\, d \bar{r}\, d\theta =\frac{\pi}{8 a^2} &
\end{flalign}
\normalsize
\noindent Please, note that the extension up to $\infty$ of the integral in $\bar r$ is a bona-fide approximation as we integrate along the long edges ($L_x, L_y$) and the exponential form of the Slater correlation factor makes the probability to be null close to the borders. We have additionally checked numerically the good performance of the approximation in the week and mid-strength confinement regime, where nanoplatelets typically lie.

\noindent Let's consider the second integral,
\small
\begin{equation}
\label{eqIII.7}
{\bf I}_2=\frac{1}{4} \iint\displaylimits_{-L/2}^{\;\;\;L/2} \cos(2 k \bar{x}_2+2 k x_1) \; e^{-2 a \,\sqrt{ \bar{x}_2^2+\bar{y}_2^2} \; } d \bar{x}_2 d \bar{y}_2 
\end{equation}
\normalsize
\noindent The identity $\cos(A+B)=\cos A \cos B- \sin A \sin B$ allows us to split ${\bf I}_2 = \frac{1}{4} ({\bf I}_{21}+{\bf I}_{22})$ with,
\small
\begin{eqnarray}
\label{eqIII.8}
{\bf I}_{21}&=&\cos 2 k x_1 \iint\displaylimits_{-L/2}^{\;\;\;L/2} \cos 2 k \bar{x}_2 \; e^{-2 a \,\sqrt{ \bar{x}_2^2+\bar{y}_2^2} \; } d \bar{x}_2 d \bar{y}_2 \nonumber \\
            &\approx & \cos 2 k x_1 \int_0^{\infty} \bar{r} d \bar{r} \; e^{-2 a \bar{r}} \int_0^{2\pi} \cos[2 k \bar{r} \cos\theta] d\theta \nonumber \\
			&=& 2 \pi \cos 2 k x_1 \int_0^{\infty} J_0(2 k \bar{r}) \; e^{-2 a \bar{r}} \; \bar{r} \, d \bar{r} \nonumber \\
			&=&   \frac{2 \pi a}{4 (a^2+k^2)^{3/2}}   \cos 2 k x_1 
\end{eqnarray}
\normalsize
\noindent where we have employed the identities:
\small
\begin{equation}
J_0(2 k \bar{r}) = \frac{1}{2 \pi} \int_0^{2 \pi} \cos[2 k \bar{r} \cos\theta] d\theta
\end{equation}
\normalsize
\noindent and 
\small
\begin{equation}
\int_0^{\infty} J_0(2 k \bar{r}) \; e^{-2 a \bar{r}} \; \bar{r} \, d \bar{r} =  \frac{a}{4 (a^2+k^2)^{3/2}}.
\end{equation}
\normalsize

\noindent On the other hand, in the integral ${\bf I}_{22}$ we meet the subintegral $\int_0^{2 \pi} \sin[2 k \bar{r} \cos\theta] d\theta = 0$. Then, ${\bf I}_{22}=0$ so that,
\small
\begin{equation}
\label{eqIII.9}
{\bf I}_2 = \frac{\pi a}{8 (a^2+k^2)^{3/2}} \;   \cos 2 k x_1 
\end{equation}
\normalsize
\noindent In an analogous way, ${\bf I}_3 = \frac{1}{4} ({\bf I}_{31}+{\bf I}_{32})$, with ${\bf I}_{32}=0$ so that, 
\small
\begin{equation}
\label{eqIII.9b}
{\bf I}_3 = \frac{\pi a}{8 (a^2+k^2)^{3/2}} \;   \cos 2 k y_1 
\end{equation}
\normalsize
\noindent Finally we deal with ${\bf I}_4$. It is convenient to write it in terms of the original $x_1, x_2$ coordinates :
\small
\begin{equation}
\label{eqIII.10}
{\bf I}_4 = \frac{1}{4} \iint\displaylimits_{-L/2}^{\;\;\; L/2}  \cos 2 k x_2  \cos 2 k y_2 \; e^{-2 a \,\sqrt{(x_2-x_1)^2+(y_2-y_1)^2}} d x_2 d y_2
\end{equation}
\normalsize
\noindent Next, we employ the identity $2 \cos A \cos B = \cos (A+B) + \cos (A-B)$ that allows to split ${\bf I}_4$ into two integrals, ${\bf I}_4= \frac{1}{8}({\bf I}_{41}+{\bf I}_{42})$, and the above change of variables: $x_2-x_1=\bar{x}_2$, $y_2-y_1=\bar{y}_2$. Then, ${\bf I}_{41}$ can be written as,

\small
\begin{eqnarray}
\label{eqIII.11}
{\bf I}_{41} &=& \iint\displaylimits_{-L/2}^{\;\;\; L/2}  \cos \left[ 2 k (\bar{x}_2 + \bar{y}_2)+ 2 k (x_1+y_1)\right] *\nonumber \\
&& \hspace{3cm} * \;\; e^{-2 a \,\sqrt{\bar{x}_2^2+\bar{y}_2^2} \;} \; d \bar{x}_2 d \bar{y}_2 \; \approx \nonumber \\
\nonumber \\
& \approx & \int_{0}^{\infty} \int_{0}^{2 \pi} \left\{ \cos\left[ 2 k \bar{r} \left( \cos\theta+\sin\theta\right)\right] \cos\left[2 k (x_1+y_1) \right] \right.- \nonumber \\
\nonumber \\
			 & & \hspace{0.5cm} - \left. \sin \left[ 2 k \bar{r} \left( \cos\theta+\sin\theta\right)\right] \sin\left[2 k (x_1+y_1) \right]\right\} *\nonumber \\
\nonumber \\			 
&& \hspace{3.5cm} *	\;\; e^{-2 a  \bar{r}} \; \bar{r} \, d \bar{r} d\theta
\end{eqnarray}
\normalsize
\noindent In the above integrals arise the subintegrals,
\small
\begin{equation}
\label{eqIII.12}
\int_0^{2 \pi} \cos\left[ 2 k \bar{r} (\cos\theta+\sin\theta)\right] d\theta = 2 \pi \, J_0(2 \sqrt{2} k \bar{r})
\end{equation}

\begin{equation}
\label{eqIII.13}
\int_0^{2 \pi} \sin\left[ 2 k \bar{r} (\cos\theta+\sin\theta)\right] d\theta = 0
\end{equation}
\normalsize

\noindent Then, 
\small
\begin{eqnarray}
\label{eqIII.14}
{\bf I}_{41} &=& 2 \pi \cos[2 k \, (x_1+y_1)] \int_0^{\infty} J_0(2 \sqrt{2} k \bar{r}) \; e^{-2 a  \bar{r}} \; \bar{r} \, d \bar{r} \nonumber \\
\nonumber \\
             &=& \frac{\pi \, a}{2 \,(a^2+ 2 k^2)^{3/2}} \; \cos[2 k (x_1+y_1)]
\end{eqnarray}
\normalsize
\noindent In a similar way we can calculate ${\bf I}_{42}$. The addition of ${\bf I}_{41}$  and ${\bf I}_{42}$ yields:
\small
\begin{equation}
\label{eqIII.15}
{\bf I}_4 = \frac{1}{8}\frac{\pi\, a}{(a^2+2 k^2)^{3/2}} \; \cos 2 k x_1 \; \cos 2 k y_1 .
\end{equation}
\normalsize

\noindent From, eqs. (\ref{eqIII.4}), (\ref{eqIII.6}),  (\ref{eqIII.9}), (\ref{eqIII.9b}) and  (\ref{eqIII.15}) we get the following closed-form for the single-particle density $\rho(r_1)$:

\small
\begin{flalign} 
\label{eqIII.16}
& \rho(r_1) = N^2 \;  \frac{\pi \, L_z}{16} \; \cos^2 \bar{k} z_1 \; \cos^2 kx_1 \; \cos^2 k y_1 \;\;* & \nonumber \\
 & \hspace{1cm} *  \left[ \frac{1}{a^2}+\frac{a}{(a^2+k^2)^{3/2}} \left(\cos 2 k x_1 + \cos 2 k y_1 \right) + \right.& \nonumber \\
 & \hspace{1cm} + \left. \frac{a}{(a^2+2 k^2)^{3/2}} \; \cos 2k x_1 \cos 2 k y_1 \right] &
\end{flalign} 
\normalsize

%............................................................................................................................%

\subsection{The single-particle density of 0D systems}
\label{subsec:6.2}

The single-particle density of particle "1" reads,

\small
\begin{flalign}
\label{eqV.2}
& \rho(r_1) = \int |\Psi(r_1,r_2)|^2 d^3r_2 = N^2 \, \cos^2 k x_1 \cos^2 k y_1 \cos^2 k z_1 * &\nonumber \\
&\hspace{1cm} *\; \iiint\displaylimits_{-L/2}^{\;\;\;L/2} d x_2\,d y_2\,d z_2\, \cos^2 k x_2 \cos^2 k y_2 \cos^2 k z_2 \;* &\nonumber \\
&\hspace{2cm} *\; e^{-2 a \,\sqrt{(x_2-x_1)^2+(y_2-y_1)^2+(z_2-z_1)^2}} &
\end{flalign}
\normalsize

\noindent As above, the identity $2 \cos^2 A = 1 + \cos 2A$ removes the squares in the cosines of the threefold integral $\bf I$ in eq. (\ref{eqV.2}) and turn it into a sum of eight integrals:
\small
\begin{equation} 
\label{eqV.3}
{\bf I}=\frac{1}{8}\left({\bf I}_1+ {\bf I}_{21}+ {\bf I}_{22}+ {\bf I}_{23}+ {\bf I}_{31}+ {\bf I}_{32}+ {\bf I}_{33}+{\bf I}_4 \right)
\end{equation} 
\normalsize
\noindent each corresponding to the different terms arising in the product of the squared cosines,
\small
\begin{flalign}
\label{eqV.4}
&\cos^2 k x_2 \cos^2 k y_2 \cos^2 k z_2 = 1+\cos 2k x_2+\cos 2k y_2 \; + &\nonumber\\ 
&\hspace{0.25cm} +\, \cos 2k z_2 + \cos 2k x_2 \cos 2k y_2 +  \cos 2k x_2 \cos 2k z_2\;+&\nonumber\\
&\hspace{0.25cm} +\, \cos 2k y_2\cos 2k z_2+\cos 2k x_2\cos 2k y_2 \cos 2k z_2. &
\end{flalign}
\normalsize 
\noindent As in previous section, in order to calculate these integrals, after changing to spherical coordinates, we extend the radial integration limit up to $\infty$ as the QD edges are large as compared to the Bohr radius of the exciton. A list of useful auxiliary integrals are collected in Appendix C.2. The different ${\bf I}_k$ integrals reads,
\small
\begin{equation} 
\label{eqV.5}
{\bf I}_1 = \iiint\displaylimits_{-\infty}^{\;\;\;\;\infty} e^{-2 a r} \; d^3r = \frac{8 \pi}{(2 a)^3}
\end{equation} 
\normalsize 
\noindent Integrals ${\bf I}_{21}, {\bf I}_{22}$ and ${\bf I}_{23}$ are similar. For example, with the notation $d\bar{v}=d\bar{x} d\bar{y} d\bar{z}$ and considering the change of variable $x_2-x_1=\bar{x}_2$, $y_2-y_1=\bar{y}_2$ $z_2-z_1=\bar{z}_2$, we have,
\small
\begin{eqnarray} 
\label{eqV.6}
{\bf I}_{23} &=& \iiint\displaylimits_{-\infty}^{\;\;\;\;\infty} \cos[2 k (\bar{z}_2+z_1)] \;\exp[-2 a \bar{r}] d\bar{v} \nonumber \\
             &=& \cos 2 k z_1 \iiint\displaylimits_{-\infty}^{\;\;\;\;\infty} \cos 2 k \bar{z}_2  \, \exp[-2 a \bar{r}] \,d\bar{v} \; -\nonumber \\
			 &&\;\;\;\; -\;   \sin 2 k z_1 \iiint\displaylimits_{-\infty}^{\;\;\;\;\infty} \sin 2 k \bar{z}_2  \, \exp[-2 a \bar{r}] \,d\bar{v} \nonumber \\
			 &=& \frac{\pi \, a}{(a^2+k^2)^2} \; \cos 2 k z_1	 
\end{eqnarray} 
\normalsize
\noindent because the integral involving the  function $\sin x$ is zero by symmetry reasons. 

\noindent Integrals ${\bf I}_{31}, {\bf I}_{32}$ and ${\bf I}_{33}$ are alike. For example, 
\small
\begin{eqnarray} 
\label{eqV.7}
{\bf I}_{31} &=& \iiint\displaylimits_{-\infty}^{\;\;\;\;\infty}  \cos[2 k (\bar{x}_2+x1)] \, 
                   \cos[2 k (\bar{y}_2+y_1)] \;  \exp[-2 a \bar{r}] d\bar{v} \nonumber \\
             &=& \cos 2 k x_1 \, \cos 2 k y_1 \iiint\displaylimits_{-\infty}^{\;\;\;\;\infty} \cos 2 k \bar{x}_2  \,\cos 2 k \bar{y}_2\,  
			     \exp[-2 a \bar{r}] \,d\bar{v} \nonumber \\
			 &-& \cos 2 k x_1 \, \sin 2 k y_1 \iiint\displaylimits_{-\infty}^{\;\;\;\;\infty}  \cos 2 k \bar{x}_2  \,\sin 2 k \bar{y}_2\,
			      \exp[-2 a \bar{r}] \,d\bar{v} \nonumber \\
			 &-& \sin 2 k x_1 \, \cos 2 k y_1 \iiint\displaylimits_{-\infty}^{\;\;\;\;\infty}  \sin 2 k \bar{x}_2  \,\cos 2 k \bar{y}_2\,
			      \exp[-2 a \bar{r}] \,d\bar{v} \nonumber \\
			 &+& \sin 2 k x_1 \, \sin 2 k y_1 \iiint\displaylimits_{-\infty}^{\;\;\;\;\infty}  \sin 2 k \bar{x}_2  \,\sin 2 k \bar{y}_2\,
			      \exp[-2 a \bar{r}] \,d\bar{v} \nonumber \\    				 
			 &=& \frac{\pi \, a}{(a^2+2\,k^2)^2} \; \cos 2 k x_1 \, \cos 2 k y_1	 
\end{eqnarray} 
\normalsize
\noindent Again, the integrals involving the  function $\sin x$ are zero by symmetry reasons.
 
\noindent Finally, out of the eight terms originated in ${\bf I}_{4}$ only the first one, that does not contain the odd function $\sin x$,  remains. With the help of integrals in Appendix C.2 we have,
\small
\begin{eqnarray} 
\label{eqV.8}
{\bf I}_{4} &=& \iiint\displaylimits_{-\infty}^{\;\;\;\;\infty}  \cos[2 k (\bar{x}_2+x1)] \, \cos[2 k (\bar{y}_2+y_1)] \; * \nonumber \\
            && \hspace{2.75cm} *   \, \cos[2 k (\bar{z}_2+z_1)]  \exp[-2 a \bar{r}]\; d\bar{v} \nonumber \\  				 
			 &=& \frac{\pi \, a}{(a^2+3\,k^2)^2} \; \cos 2 k x_1 \, \cos 2 k y_1 \, \cos 2 k z_1	 
\end{eqnarray}                                                                                            
\normalsize 
\noindent From eqs. (\ref{eqV.2},\ref{eqV.3},\ref{eqV.5}-\ref{eqV.8}) we obtain the following closed-form for the single-particle density: 
\small
\begin{flalign}
\label{eqV.9}
& \rho(r_1) = N^2 \, \frac{\pi}{8} \; \cos^2 k x_1 \; \cos^2 k y_1 \; \cos^2 k z_1  \;* & \nonumber \\
&\hspace{0.5cm} * \;  \left\{ \frac{1}{a^3}+\frac{a}{(a^2+k^2)^2} \left(\cos 2 k x_1  + \cos 2 k y_1  +\cos 2 k z_1 \right) 
              \right. & \nonumber   \\
&\hspace{1cm}  + \frac{a}{(a^2+2k^2)^2} \left[ \cos 2 k x_1 \, \cos 2 k y_1  \; + \right. & \nonumber   \\
&\hspace{2.5cm} \left. +\cos 2 k x_1 \, \cos 2 k z_1 + \, \cos 2 k y_1 \, \cos 2 k z_1 \right] \nonumber   \\
&\hspace{1cm}  \left. + \frac{a}{(a^2+3k^2)^2} \; \cos 2 k x_1 \, \cos 2 k y_1  \, \cos 2 k z_1 \right\}&
\end{flalign}
\normalsize  

%............................................................................................................................%
\section{Some useful integrals}
\label{section:7}
\subsection{Useful integrals for the quasi-1D model}
\label{subsection:7.1}

\small
\begin{flalign} 
&\int_{-L/2}^{L/2} dx \cos^2 (\frac{\pi}{L} x) =  \frac{L}{2} &\\
&\int_{-L/2}^{L/2} dx \cos^3 (\frac{\pi}{L} x) = \frac{4 L}{3 \pi} &\\
&\int_{-L/2}^{L/2} dx \cos^4 (\frac{\pi}{L} x) = \frac{3 L}{8} &  \\
&\int_{-L/2}^{L/2} dx \cos^2 (\frac{\pi}{L} x) \cos (2\;\frac{\pi}{L} x) =  \frac{L}{4} &\\
& \int_{-\infty}^{\infty} d\bar{x}_2 \cos^2 k(\bar{x}_2+x_1) \; e[{-2 a |\bar{x}_2|} = &\nonumber\\
&\hspace{3cm} = \,  \frac{1}{2a}+\frac{1}{2} \frac{a}{a^2+k^2} \; \cos 2 k x_1 &\\
&\int_{-\infty}^{\infty} d\bar{x}_2 \cos k(\bar{x}_2+x_1) \; e^{- a |\bar{x}_2|} = \frac{2 a}{a^2+k^2} \; \cos k x_1 &\\
&\int_{-\infty}^{\infty} d\bar{x} \; \sin[2 k(\bar{x}+x_1)]\,\frac{|\bar{x}|}{\bar{x}} \; e^{-2 a |\bar{x}|} = &\nonumber\\
&\hspace{4cm} = \frac{k}{a^2+k^2} \; \cos 2k x_1 & \\
& \int_{-\infty}^{\infty} dx_2 \; \cos^2 k x_2 \; e^{-2 a |x_2-x_1|} = & \nonumber \\
&\hspace{1cm} = \int_{-\infty}^{\infty} d\bar{x}_2 \; \cos^2 k \,(\bar{x}_2+x_1) \; e[{-2 a |\bar{x}_2|} \,\, & \nonumber \\
&\hspace{1cm} =\int_{-\infty}^{\infty} d\bar{x}_2 \; \left(\frac{1+\cos 2 k\,(\bar{x}_2+x_1)}{2} \right) \; e^{-2 a |\bar{x}_2|} & \nonumber\\
&\hspace{1cm} = \frac{1}{2 a} +\frac{a}{2\, (a^2+k^2)} \; \cos 2 k x_1
\end{flalign}
\normalsize 

\subsection{Useful integrals for the 0D model}
\label{subsection:7.2}
In the following integrals $\xi$ represents the coordinate $x$, $y$ and $z$. $\xi_1 \, \xi_2$ represents the product two different coordinates and $dv=dx dy dz$:
\small
\begin{flalign} 
&\iiint\displaylimits_{-\infty}^{\;\;\;\;\infty} dv \; e^{-a r} = \frac{8 \pi}{a^3} &\\
&\iiint\displaylimits_{-\infty}^{\;\;\;\;\infty} dv \, \cos k\xi \; e^{-a r} = \frac{8 a \pi}{(a^2+k^2)^2}& \\ 
&\iiint\displaylimits_{-\infty}^{\;\;\;\;\infty} dv \, \cos k\xi_1 \, \cos k\xi_2 \; e^{-a r} = \frac{8 a \pi}{(a^2+2 k^2)^2} &\\
&\iiint\displaylimits_{-\infty}^{\;\;\;\;\infty} dv \, \cos k x \, \cos k y \, \cos kz \; e^{-a r} = \frac{8 a \pi}{(a^2+3 k^2)^2} &\\
&\iiint\displaylimits_{-\infty}^{\;\;\;\;\infty} dv \, \cos k_1 \xi_1 \, \cos k_2 \xi_2 \; e^{-a r} = \frac{8 a \pi}{(a^2+k_1^2+k_2^2)^2} &\\
& \iiint\displaylimits_{-\infty}^{\;\;\;\;\infty} dv  \, \cos k_1 x \, \cos k_2 y \, \cos k_3 z \; e^{-a r} = &\nonumber\\
&\hspace{3.5cm} =  \; \frac{8 a \pi}{(a^2+k_1^2+k_2^2+k_3^2)^2} &   
\end{flalign}  
\normalsize 

%%%%%%%%%%%%%%%%%%%%%%%%%%%%%%%%%%%%%%%%%%%%%%%%%%%%%%%%%%%%%%%%%%%%%%%%%%%%%%%%%%%
\section{Reducing the integral multiplicity} 
\label{section:8}

The basic idea in the reduction of a twofold  into a single numerical integration is to use a set of variables allowing for an analytical integration of one of the two variables. In our case, we deal with:
\small
\begin{equation}
\label{Ap4.1}
\iint\displaylimits_{-L/2}^{\;\;\;\;L/2} dx_e dx_h \cos^2 k x_e \, \cos^2 k x_h \; f(|x_e-x_h|) 
\end{equation}  
\normalsize 

\noindent After the cosmetic change of variables $k x_i=\xi_i$, turning the integration original limits $[-L/2,L/2]$ into $[-\pi/2,$ $\pi/2]$, we define $\xi =  \xi_e-\xi_h$ and $\xi' =  \xi_e + \xi_h$. This transformation has a $1/2$ Jacobian, i.e., $dx_e dx_h  \to \frac{1}{2} d\xi d\xi'$, and turns the rectangular integration region into a rhomboidal one with the vertices of the rhombus separated a distance $\pi$ from the origin.

\noindent It should be said that a similar transformation can be employed to calculate the Coulomb integrals in the SCF calculation (section II), as we met the integral:
\small
\begin{equation}
\label{Ap4.2}
\iint\displaylimits_{0}^{\;\;\;\;2 \pi} \frac{d\theta_1\, d\theta_2}{\sqrt{r_1^2+r_2^2 - 2 r_1 r_2 \cos(\theta_2-\theta_1)} }
\end{equation} 
\normalsize 
\noindent In this case, integrand periodicity allows either to integrate in a rhombus with vertices $[(-2\pi,2\pi),$ $(0, 4\pi),$ $(2\pi,2\pi),$ $(0,0)]$ or in a rectangle
with $\theta\in [0,2\pi]$ and $\theta' \in [0,4\pi]$, the second option disentangling the integration of either coordinate. The integration on $\theta'$ just yields $4 \pi$, so that the above twofold integral turns into a single coordinate integral, 
\small
\begin{equation}
\label{Ap4.3}
\int_{0}^{2 \pi} \frac{d\theta}{\sqrt{r_1^2+r_2^2 - 2 r_1 r_2 \cos\theta} }.
\end{equation}
\normalsize 
\noindent All the same, the integrand in eq.  (\ref{Ap4.1}) is not periodic. This is due to the fact that $f(|x_e-x_h|)$ represents the exponential function divided by the modulus of the electron-hole distance. Then, we must integrate the rhombus region. We do it by analytically integrating over $\xi'$, keeping $\xi$ constant, between limits $[\xi-\pi,\pi-\xi]$ and then numerically integrating $\xi$ within the limits $[0,\pi]$. This actually corresponds to one half of the rhombus region, as $\xi \in[-\pi,\pi]$. However, for symmetry reasons,  both half regions integration yield the same result so the required integral is just twice the one calculated.

\noindent  Finally, it should be pointed out that should we enclose the Coulomb polarization, i.e., include all image charges originated by the dielectric mismatch, then, instead of an integrals of squared cosines times $f(|\xi|)$, we  have a large sum of inverse of squared roots including  both $\xi =  \xi_e-\xi_h$ and $\xi' =  \xi_e + \xi_h$, so that the integrals over $\xi'$ cannot be done analytically.



\end{document}